\documentstyle[preprint,aps]{revtex}
\begin{document}
\draft
\title{\bf{\LARGE{A Study of The Formation of Stationary 
Localized States Due to
Nonlinear Impurities Using The Discrete Nonlinear Schr$\ddot{o}$dinger
Equation}}}
\author{B. C. Gupta, K. Kundu}
\address{Institute of physics, Bhubaneswar - 751 005, India}
\maketitle
\begin{abstract}
The Discrete Nonlinear Schr$\ddot{o}$dinger Equation is used to 
study the formation of stationary localized states due to a single 
nonlinear impurity in a Caley tree and a dimeric nonlinear impurity 
in the one dimensional system. The rotational nonlinear impurity and
the impurity of the form $-\chi \mid C \mid^{\sigma}$ where $\sigma$ 
is arbitrary and $\chi$ is the nonlinearity parameter are considered.
Furthermore, $\mid C \mid$ represents the absolute value of the 
amplitude. Altogether four cases are studies. The usual Greens 
function approach and the ansatz approach are coherently blended 
to obtain phase diagrams showing regions of different number of 
states in the parameter space. Equations of critical lines separating 
various regions in phase diagrams are derived analytically.
For the dimeric problem with the impurity 
$-\chi \mid C \mid^{\sigma}$, three values of $\mid \chi_{cr} \mid$, 
namely, $\mid \chi_{cr} \mid = 2$, at $\sigma = 0$ and $\mid \chi_{cr} 
\mid = 1$ and 
$\frac{8}{3}$ for $\sigma = 2$ are obtained. Last two values are 
lower than the existing values. Energy of the states as 
a function of parameters is also obtained. A model derivation for 
the impurities is presented. The implication of our results in 
relation to disordered systems comprising of nonlinear impurities 
and perfect sites is discussed.
\end{abstract}
\narrowtext
\newpage
\section{Introduction}

The Discrete Nonlinear Schr$\ddot{o}$dinger equation (DNLS)
is a set of N coupled differential equations.
\begin{eqnarray}
i \frac{dC_{m}}{dt} &=& - \chi_{m} f_{m}(\mid C_{m} \mid) C_{m} 
+ V_{m,m+1} C_{m+1} + V_{m,m-1} C_{m-1} \nonumber \\
{\rm where}~~~~~~ V_{m,m+1} &=& V_{m+1,m}^{\star};~~ {\rm and}
~~m=1,2,3,.........N.
\end{eqnarray}
In eq.(1) the nonlinearity appears through functions $f_{m}(\mid C_m
\mid)$ and $\chi_{m}$ is the nonlinearity parameter associated with
the m-th grid point. Since, $\sum_{m} \mid C_m \mid^2$ is made to
unity by choosing appropriate initial conditions, $\mid C_m \mid^2$
can be interpreted as a probability of finding particle at the m-th
grid point. One way to derive this set of equations is to couple in
the adiabatic approximation the vibration of masses at the lattice
points of a lattice of N sites to the motion of a quasi particle in
the same lattice. The motion of the quasi particle is described,
however, in the frame work of a tight binding Hamiltonian (TBH).
Same type of equation can also be obtained by nonlinear coupling of
anharmonic oscillators through both positions and momenta of the 
oscillators \cite{1,2,3,4,5,6,7,8,9,10,11,12,13}. The set of
equations, thus derived, are called the discrete self--trapping  
equation (DST). These equations also posses a constant of motion 
analogous to $\sum_{m} \mid C_m \mid^2$ in the DNLSE. In fact, 
both the DST and the DNLSE contain the same number of constants of 
motion.

The analytical solution of eq.(1) in general are not known.
However, for nonlinear quantum dimers which are two sites systems with
the nonlinearity either on both the site--energies or in one of them can
be solved analytically for any arbitrary initial condition. From
analytical solutions a self tapping transition is found in this
model. The trapping of hydrogen ions around oxygen atoms in metal
hydrides and the energy transport from the absorption center to the
reaction center in photosynthetic 
unit have been modeled by the effective quantum nonlinear dimer 
\cite{14,15,16,17,18,19,20,21,22,23,24,25,26}. The nonlinear dimer 
analysis has also been applied to several experimental situations, 
like the neutron scattering of hydrogen atoms trapped at impurity 
sites \cite{27,28}.

The self--trapping of a quasi particle at the nonlinear site in a
perfect lattice containing a single nonlinear impurity has been
studied. In the analysis $f(\mid C_m \mid)$ has been taken to be
$\mid C_m \mid^2$. The critical value of the nonlinear parameter,
$\chi$ for the self trapping is found to increase with increasing the
dimension \cite{29}. It has also been shown that near the self--trapping
transition the dynamics of the quasiparticle is mostly confined to
the few near neighbors \cite{30}. An interesting experimental example in this
context is the observation that trapped hydrogen atoms in metals like
Nb move among the sites in the neighborhood of impurity atoms such as
oxygen.\cite{15,27,28} Self trapping transition has
also been obtained in systems where a nonlinear cluster is embedded in
a perfect lattice. In fact, two types of transitions are obtained,
with cluster--trapping transition preceding the self--trapping
transition \cite{31}. 

Another important feature of this type of nonlinear equations is that
these can yield stationary localized ( SL ) states and soliton--like
solutions. However, the presence of disorder or aperiodicity 
either in site energies or in nonlinearity parameters is needed 
for this purpose. For example, the Ablowitz--Ladik like equation 
\cite{32,33,34,35,36,37,38} 
can yield both localized and soliton like solution in the presence 
of disorder in static site energies. Similarly, the DNLSE can 
produce soliton--like solution if any aperiodicity is present in the
static site energies \cite{39}. The effect of disorder in the static site
energies, in the nonlinear parameters and in both of these on the solution 
of the DNLSE has also been studied by numerical integrations. One also
finds soliton--like solutions in this case. It has also been shown
that the presence of a nonlinear impurities in the absence of any disorder
in the static site energies can produce SL states
in one dimension.
The first study was made using Greens function approach and authors
considered $f(\mid C_m \mid)$ = $\mid C_m \mid^2$ \cite{40}. Furthermore,
some cursory discussion on the two and more nonlinear impurities has
been presented. Later, the one nonlinear impurity case has been
generalized by taking $f(\mid C_m \mid) = \mid C_m \mid^{\sigma}$ and
formation of SL states is studied in one, two and three
dimensions \cite{41,42,43}.

It is obvious from the above discussion that for a complete
understanding of the single nonlinear impurity problem the effect 
of increasing of connections on the formation of SL states in the
DNLSE needs further consideration. So, we plan to study the formation of
SL states in Caley tree in the presence of a single nonlinear 
impurity. Along with $f(\mid C_m \mid)$ = $\mid C_m \mid^{\sigma}$,
the effect of more generalized nonlinearity impurity ( like
rotational nonlinear impurity ) needs also
consideration. Furthermore, a thorough understanding of the nonlinear 
dimer problem in relation to SL states is required. 
This is a part requirement to understand 
the effect of disorder in nonlinearity parameters in the solutions of
the DNLSE. So, we study here in substantial detail the phase diagram 
of SL states due to the presence of a nonlinear dimer
in the perfect one dimensional chain. In this connection we also
synthesize the original Green function approach and the
ansatz approach \cite{40,41}. Finally, we present a logical 
derivation of DNLSE with $f(\mid C_m\mid)$ = $\mid C_{m}\mid^
{\sigma}$ by generalizing the potential of the nonlinear pendulum. 
To the best of our knowledge no such attempt has been made 
in this direction. 

The organization of the paper is as follows.In the forthcoming 
section we consider a nonlinear impurity of the
form $-\chi \mid C \mid^{\sigma}$ embedded in a Caley tree. We next
consider in sec.III, a rotational nonlinear impurity in the Caley
tree. In sec.IV we deal with a dimeric nonlinear impurity in an one
dimensional system. The impurity is of course, of the form $-\chi
\mid C \mid^{\sigma}$. A similar problem with the rotational
nonlinear impurity is considered in sec.V. The penultimate section
deals with a model derivation of the DNLSE considered here. Finally
we conclude our paper by highlighting the major features and the
importance of our results in understanding the behavior of disordered
systems made up of nonlinear impurities and perfect sites.

\section{A single nonlinear impurity of the form $-\chi 
\mid C \mid^{\sigma}$ in a Caley tree}

We consider a Caley tree with the coordination number, Z. Its
connectivity is, therefore K = Z-1. We put a nonlinear impurity at a
site and call it the zeroth site. Furthermore, we consider here
$f_{0}(\mid C_0 \mid)$ = $\mid C_0 \mid^{\sigma}$ where $\sigma \ge
0$. Now, from this site Z branches will emerge. 
Again from each of these branches K
branches will come out. So, the number of points in the n-th
generation (or n-th shell) is $Z K^{(n-1)}$. From any point in the n-th
shell a quasiparticle can move either to K points in the (n+1)-th shell
or to a point in the (n-1)-th shell if n$\ge$ 1. Of course, we are
considering only the nearest neighbor hopping with real hopping
element. So, if the nearest neighbor hopping element is same for 
all sites, every point in a given shell
will show the identical time evolution of the probability amplitude
of the quasiparticle. Thus, the set of equations that describes the time
evolution of the probability amplitude in the present model is
\cite{44}.
\begin{equation}
i\frac{dC_n}{dt} = K C_{n+1} + C_{n-1},~~~~~~~~~~~n\ge 1
\end{equation}
\begin{equation}
i \frac{dC_0}{dt} = -\widetilde \chi \mid C_{0} \mid^{\sigma} C_{0} + Z
C_{1},~~~~~n=0 
\end{equation}
Furthermore, we have
\begin{equation}
\mid C_{0}(t) \mid^2 + \frac{Z}{K} ~\sum_{n=1}^{\infty} K^n \mid
C_{n} \mid^2 = 1
\end{equation}
Since we are interested here in SL states we write 
$C_{n}(t) = \exp(-i \tilde E t) a_{n}$. After introducing this form
of $C_{n}(t)$ in eq.(2) and eq.(3) we make the following
transformations: (i) $a_{n} = K^{-n/2} \phi_{n}$, (ii) E = $\tilde E
/\sqrt{K}$ and (iii) $\chi = \tilde \chi / \sqrt{K}$. With this
transformations we finally obtain 
\begin{eqnarray}
{ E \phi_n} &=& {\phi_{n+1} +
\phi_{n-1},~~~~~~~~~~~~~~~~~~~~~ n \ge 1} \\
{E \phi_0} &=& {-\chi \mid \phi_0 \mid^{\sigma} \phi_0 + \frac{Z}{K} 
~\phi_1,~~~~~~~~~~n=0} \\
{\rm and}~~~~~~~~~~~~~~~~~~~~~~~~~~~~~ 
\mid \phi_0 \mid^2 &+& \frac{Z}{K} ~\sum_{n=1}^{\infty} \mid \phi_n
\mid^2 = 1
\end{eqnarray}
After some standard algebra for $\mid E \mid > 2$ \cite{32} we obtain 
from eq.(5) and eq.(6)
\begin{equation}
\frac{\phi_n}{\phi_0} = \frac{E - \chi K \mid \phi_0 \mid^{\sigma}}{Z}
G_{n,0}^0(E) - G_{n+1,0}^0(E)
\end{equation}
\noindent where $G_{n,0}(E) = [sgn(E)]^{n+1}$ $G_{0,0}^0(\mid E 
\mid)$ $\eta^{n}$ and $sgn(E)$ denotes the signature of $E$. We 
also have   $\eta = \frac{\mid E \mid - \sqrt{E^2 - 4}}{2}$. So, 
$0 \le \eta \le 1$. For $G_{0,0}^0(\mid E \mid)$ we have 
\begin{equation}
G_{0,0}(\mid E \mid) = \frac{1}{\sqrt{E^2 - 4}} = \frac{\eta}{1 - \eta^2}
\end{equation}
Eq.(8) for $n$ = 0 yields
\begin{equation}
1 - \frac{E G_{0,0}^0(E)}{Z} + G_{1,0}^0(E) = -\frac{\chi K}{Z} \mid \phi_0
\mid^{\sigma} G_{0,0}^0(E)
\end{equation}
So, the energy of SL states is obtained from solution of
eq.(10).  We further note that the left hand side of eq.(10) is
independent of the sign of $E$. On the other hand, to keep the sign
unchanged in the right hand side of eq.(10) a simultaneous
change of the sign of $E$ and $\chi$ is required. Eq.(10) when written
in terms of $\eta$ assumes the form 
\begin{equation}
\frac{K}{\eta} - \eta = -\chi K \mid \phi_0 \mid^{\sigma} sgn(E)
\end{equation}
Similarly eq.(8) after some simple algebra yields for $n \ge 1$ 
\begin{equation}
\phi_{n} = \phi_0 ~\eta^{n}
\end{equation}
Now introducing eq.(12) in eq.(7) we obtain 
\begin{equation}
\eta^2 = \frac{K(1 - \mid \phi_{0} \mid^2)}{K + \mid \phi_0 \mid^2}
\end{equation}
Using eq.(13) we eliminate $\eta$ from eq.(11) to obtain the equation
for $\mid \phi_0 \mid$. This in turn yields
\begin{equation}
\frac{4}{\widetilde \chi^2 } = \mid \phi_0 \mid^{2 \sigma} 
\frac{(K + 1)^2 - ((K -1) + 2 \mid \phi_0 \mid^2)^2}{(K - 1 + 
2 \mid \phi_0 \mid^2)^2} 
\end{equation}
For $K$ = 1, ~equation (14) then gives 
\begin{equation}
\frac{4}{\chi^2} = \mid \phi_0 \mid^{2 (\sigma - 2)} (1 - \mid \phi_0
\mid^4) 
\end{equation}
and this equation has already been obtained \cite{41}. We again note
that $\mid \phi_0 \mid^2 \le 1$. The energy ($E$) of SL
states is
\begin{equation}
E = -sgn(\chi) ~(\eta + \frac{1}{\eta}) = -sgn(\chi) \frac{(2 K - (K
-1) \mid \phi_0 \mid^2)^2}{\sqrt{K} \sqrt{(K + 1)^2 - ((K - 1) + \mid
\phi_0 \mid^2)^2}}
\end{equation}
Equation (16) can be derived using eq.(13) and eq.(14). We
note that for $\sigma$ = 0, eq. (16) in conjunction with eq.(14)
yields
\begin{equation}
E = -sgn(\chi) ~[ -\frac{\mid \chi \mid (K - 1)}{2} + \frac{K + 1}{2
\sqrt{K}} \sqrt{\chi^2 K + 4} ]
\end{equation}
Eq.(17) yields the correct result for a single linear impurity
in an one dimensional perfect system. We also see from eq.(16) that
for $K = 1, ~E = -sgn(\chi) \mid \widetilde \chi \mid \mid \phi_0
\mid^{\sigma - 2}$. Then, for $\sigma = 2, ~\mid E \mid = \mid
\widetilde \chi \mid$. Inasmuch as the minimum absolute value of the 
energy of a SL state is $2+$, the critical value of $\mid  
\chi \mid ~i.e., \mid \widetilde \chi_{cr} \mid = 2$. Above this 
value of $\mid \widetilde \chi \mid$, two SL states are obtained 
\cite{41}.

We numerically solve eq.(14) for $\mid \phi_{0} \mid$ as a function 
of $\chi$ and $\sigma$. If $\mid \phi_0 \mid \le 1$, we obtain an
admissible solution. Number of such solutions are the number of
SL states in the system. We then calculate the 
energy of the SL states using eq.(16). The phase diagram 
for SL states and energies of these states as a function 
of $\chi$ and $\sigma$ are shown in Fig.1 and Fig.2 for K=1, 2, 3, 10
respectively. We note that 
for $\sigma = 0$ we have the standard Anderson localization problem. 
To obtain a localized eigenstate in a Caley tree of connectivity K, 
the condition $\mid \widetilde \chi_{cr} \mid \ge \frac{K - 1}{\sqrt
{K}}$ has to be satisfied. This can be obtained from eq.(14) by
setting $\mid \phi_0 \mid^2 = 0$ and also from equation (17) by 
setting $\mid E \mid = 2$. This is also obtained in our numerical 
analysis as shown in the Fig.1. For K = 1, there are three regimes. 
When $\sigma \le 2$, there is always one SL state for $\mid 
\chi \mid > 0$. For $\sigma \ge 2$, there is a critical value of $
\chi$ ($\mid \chi_{cr} \mid$) for each $\sigma$ ( represented by 
solid curve ) below which no SL state appears but above it two SL 
states ensue. On the other hand, for $K > 1$, we find that for 
any value of $\sigma > 0$, there is a $\mid \chi_{cr} \mid$ below which no 
SL state appears in contrary to the $K = 1$ case.
But for $\mid \chi \mid > \mid \chi_{cr} \mid$, we get two SL states.
So, there is a critical line ( dashed line for K = 2, dotted line for
K = 3, dot-dashed line for K = 10 in Fig.1. ) in the ( $\chi$, $\sigma$ ) 
plane which separates the no SL state region and the region containing
two SL states. If one approaches  the critical line from above for
any $\sigma$, two distinct SL states merge on the
critical line and below that line they disappear. As we
increase K, the separation between critical lines for any $\sigma$
decreases as shown in Fig.1.  So, $\mid \chi_{cr} \mid$
becomes virtually independent of the connectivity for very large K.
For the energy diagram ( see Fig.2 ) we have taken $\sigma = 1$. 
We see that for $K = 1$ the energy of the SL state increases with
$\chi$. This implies that localization becomes stronger for larger
$\mid \chi \mid$. On the other hand for $K = 2, 3, 10$ {\i.e.}, for $K > 1$,
one of the two SL states increases with $\mid \chi \mid$. This corresponds to
strong localization for large $\mid \chi \mid$. The energy of the other SL 
state goes towards the appropriate band edge energy. This corresponds to 
weak localization for large $\mid \chi \mid$.

We now present analytical results in favour of our numerical results. The
relevant equation for this purpose is eq.(14). We denote right hand
side expression of eq.(14) by $g_{K,\sigma}(A)$, where $A = \mid \phi_0
\mid^2$. We note that for $\sigma > 0$, ~$g_{K,\sigma}(0) = 0 = 
g_{K,\sigma}(1).$ We also have ~$g_{K,\sigma}(A) \ge 0$ for A in
[0,1]. So, there must be a maximum
of this function with respect to A for A in [0,1]. For $\sigma = 0$,
this maximum occurs at A = 0. If the value of left hand side 
of eq.(14) lies above the maximum, no SL states will be obtained.
However, for ($4/\mid \widetilde \chi \mid^2$) lying below the maximum
we shall obtain at least two SL states. Our
numerical result suggests that $g_{K,\sigma}(A)$ has one maximum for
A in [0,1]. So, by imposing the maximality condition on
$g_{K,\sigma}(A)$ with respect to A {\em i.e.}, by setting $\partial
g_{K,\sigma}(A) / \partial A = 0$, we obtain 
\begin{equation}
2 \sigma A^3 + 3 \sigma (K - 1) A^2 + [(K + 1)^2 + (K^2 - 4 K + 1)
\sigma] A - \sigma K (K - 1) = 0
\end{equation}
Note that for $\sigma = 0$, A = 0. Furthermore, for K = 1, we get $A
= \sqrt{(\sigma - 2)/\sigma}$. For $K > 1$ and $\sigma > 0$ eq.(18)
has only one real root, RR given by
\begin{eqnarray}
RR &=& \frac{(1 - K)}{2} + \frac{3^{2/3} (\sigma -2) (K + 1)^2 +
3^{1/3} D^2}{6 \sqrt{\sigma} D} \\
{\rm where,}~~~~~~~~~~~~~ D &=& [9 \sqrt{\sigma} (K + 1)^2 (K - 1) + 
\sqrt{3} (K + 1)^2
D_{0}]^{1/3}\\
{\rm and}~~~~~~~~~~~~~~~ D_{0} &=& \sqrt{(K + 1)^2 (2 - \sigma)^3 + 27 \sigma
(K - 1)^2}
\end{eqnarray} 
\noindent We emphasize that for all value of $\sigma > 0$ and $K >
1$, RR is real and lies between 0 and 1. So, the equation for the
critical line in the $(\chi, \sigma)$ plane separating the no state
and the two states regions is
\begin{equation}
\mid \chi_{cr} \mid = \frac{2}{\sqrt{K} \sqrt{g_{k,\sigma}(RR)}}
\end{equation}
We now analyze the case of K = 1 along the same line. The relevant
equation is eq.(15) in which the right hand side will be denoted by
$g_{\sigma}(A)$. Inasmuch as $0 \le A \le 1$ the domain of
$g_{\sigma}(A)$ is $[ 0, \infty ]$ if $\sigma < 2$. So, for $\sigma <
2$ the line y = constant = $(4/\chi^2)$ always intersect the curve
$g_{\sigma}(A)$ at a single point. This will happen even if $\mid \chi
\mid$ is infinitesimally small. So, for $\sigma < 2$ we shall always
get one and only one SL state if $\mid \chi \mid >
0$. Precisely this is obtained in \cite{41}. If $\sigma > 2, ~g_{\sigma}(0)
= 0 = g_{\sigma}(1)$. Furthermore, $g_{\sigma}(A)$ is positive
semidefinite for A in [0, 1]. So, there must be at least one maximum
for A in [0, 1]. We also note that for $\sigma = 2$, the
maximum of $g_{\sigma}(A)$ occurs at A = 0. Therefore, for K = 1 and
$\sigma = 0$, applying same analysis we obtain $\mid \chi_{cr} \mid 
= 2$ below which there is no SL state and above it there is
one SL state. Furthermore, it can be shown from eq.(15) by maximizing
$g_{\sigma}(A)$ that $\mid \chi_{cr} \mid$ separating the no state 
and two state regions for $\sigma > 2$ is
\begin{equation}
\mid \chi_{cr} \mid = \frac{\sqrt{2} \sigma^{\sigma /4}}{(\sigma -
2)^{(\sigma - 2)/4}}
\end{equation}
This expression for $\mid \chi_{cr} \mid$ has been mentioned \cite{41} 
without any derivation.

\section{A single rotational nonlinear impurity in a Caley tree}

We consider here the case where $f(\mid C_0 \mid) = \frac{\mid C_0
\mid}{\sqrt{1 + (\frac{\chi}{\Delta})^2 \mid C_0 \mid^2}}$. The
origin of this particular form of nonlinearity has been discussed in
\cite{45}. For this case from the appropriate version of eq.(10) we
obtain for $\mid \phi_0 \mid^2$ = A
\begin{eqnarray}
A^4 &+& (K - 1) A^3 + P A^2 + Q = 0 \\
{\rm where,}~~~~~~~~~~~~~
P &=& \frac{4 - \chi^2 (K^2 - ( \frac{K - 1}{\Delta})^2)}{\chi^2
(\frac{4}{\Delta ^2} + K)} \\
{\rm and}~~~~~~~~~~~~~~~
Q &=& (K - 1)^2 / [\chi^2 ( \frac{4}{\Delta ^2} + K)]
\end{eqnarray}
For K=1, that is for a one dimensional system, we find from eq.(24)
\begin{equation}
A = \frac{\sqrt{\chi^2 - 4}}{\sqrt{\chi^2 + \Delta^2}} \mid
\frac{\Delta}{\chi} \mid
\end{equation}
So if $\mid \chi \mid > 2$, one and only one SL  
state will be obtained. Furthermore, the energy ($E$) of this 
SL state is
\begin{equation}
E = \pm \mid \chi \mid ~\sqrt{\frac{4 + \Delta ^2}{\chi^2 + \Delta ^2}}
\end{equation}
However, $K > 1$, the analytical calculation of the energy of the
SL state is quite prohibitory. So, we examine 
the case for $K = 2$. Our results are shown in Fig.3. We find 
again two regions. In the lower region we find no SL
state. On the other hand in the upper region
there are two SL states. The line separating 
these regions contains one SL state. We find 
that for $\mid \Delta \mid \rightarrow 0$ , $\mid \chi_{cr} \mid 
\rightarrow \infty$. This should be expected.
Another interesting feature is that $\mid \chi_{cr} \mid$ reduces
with increasing $\mid \Delta \mid$. For small value of $\mid \Delta
\mid$, the fall in $\mid \chi_{cr} \mid$ is quite sharp with
increasing $\mid \Delta \mid$.

\section{A nonlinear dimer impurity in an one--dimensional perfect
chain: Impurity of $-\chi \mid C \mid^{\sigma} type$} 

Two consecutive sites occupied by nonlinear impurities are labeled
as zeroth site and one-th site respectively. We consider here,
$\chi_{0} ~f_{0}(\mid C_{0}) \mid) = \chi \mid C_0 \mid^{\sigma}$ and
$\chi_1 ~f(\mid C_1 \mid) = \chi \mid C_1 \mid^{\sigma}$. All other
${\chi_m}$ are set to zero. All ${V_{n,n+1}}$ in eq.(1) are set
to unity. Since we are looking for SL states, we set $C_n(t) =
\phi_n \exp{(-iEt)}$ in the appropriate version of eq.(1). From
the set of equations thus obtained we solve for ${\phi_n}$ by
the standard procedure \cite{32}. Consequently, we obtain
\begin{equation}
\phi_n = G_{n,0}^0(E) ~\epsilon_0 ~\phi_0 + G_{n,1}^0(E) ~\epsilon_1
~\phi_1
\end{equation}
In equation (29) for the purpose of generality we define $\epsilon_0
= -\chi \mid \phi_0 \mid^{\sigma}$ and $\epsilon_1 = -\chi \mid
\phi_1 \mid^{\sigma}$. Furthermore, for $n$ = 0, and $n$ = 1 eq.(29)
yields  
\begin{eqnarray}
\phi_0 &=& G_{0,0}^0(E) ~\epsilon_0 ~\phi_0 + G_{0,1}^0(E) ~
\epsilon_1 ~\phi_1\\
{\rm and}~~~~~~~~~~~~~
\phi_1 &=& G_{1,0}^0(E) ~\epsilon_0 ~\phi_0 + G_{1,1}^0(E) ~
\epsilon_1 ~\phi_1
\end{eqnarray}
We note that $G_{n,m}^0(\mid E \mid)$ has been defined by equation
(9). After some simple algebra it can be shown with the help of
eq.(30) and eq.(31) that 
\begin{eqnarray}
\phi_n = (sgn(E) ~\eta)^{n-1} ~\phi_1, ~~~~~~~~~~~~~~&n& > 1 \\
\phi_{-\mid n \mid} = (sgn(E) ~\eta)^{\mid n \mid}
~\phi_0,~~~~~~~~~~~~&\mid n \mid& > 0
\end{eqnarray}
where $0 \le \eta \le 1$ and it is defined in the section before. 
We further note that the form of ${\phi_n}$ given by eq.(32) 
and eq.(33) is independent of the form of $\epsilon_0$ and
$\epsilon_1$. Since we must have $\sum_{-\infty}^{\infty} \mid 
\phi_n \mid^2 = 1$, using eq.(32) and eq.(33) we find that 
\begin{equation}
\mid \phi_0 \mid^2 + \mid \phi_1 \mid^2 = 1 - \eta^2
\end{equation}
\noindent When $\epsilon_0$ and $\epsilon_1$ are linear impurities, from
eq.(30) and eq.(31) we obtain 
\begin{equation}
E_{\pm} = \frac{\epsilon_0 \epsilon_1 ( \epsilon_0 + \epsilon_1 ) ~\pm
\mid 2 - \epsilon_0 \epsilon_1 \mid \sqrt{( \epsilon_0 - \epsilon_1
)^2 + 4}}{2 ~( \epsilon_0 ~\epsilon_1 - 1 )}
\end{equation}
\noindent where $E_{\pm}$ denotes the energy of 
the SL state. We note
that for $\epsilon_0 = \epsilon_1 > 2$, two SL states appear
above the band. The situation reverses for $\epsilon_0 = \epsilon_1 <
-2$. Furthermore for $\epsilon_0 = \epsilon_1 \rightarrow 0$, we get
$E_{\pm} = \pm 2$. In other words, we get one of the band edge states.
These are consistent with established results \cite{46} but the exact
form of the solution is not presented there. In
the present model we have $\epsilon_0 = \epsilon_1 = -\chi$ for
$\sigma = 0$. On the other hand, since $\mid \phi_0 \mid, \mid \phi_1
\mid < 1$, for a finite $\chi$, if $\sigma \rightarrow \infty$, we get
$\epsilon_0 = \epsilon_1 \rightarrow 0$.
General solutions of the nonlinear dimer impurity
problem are expected to show these asymptotic behaviors. 
It should be noted in this context that for a single nonlinear 
impurity this continuity has
been rigorously established \cite{41}. 

We now consider the case of our nonlinear dimer. We have three
equations, namely, eq.(30) and eq.(31) and eq.(34). So, if $\phi_0$
and $\phi_1$ are fully complex quantities, for the problem to be well
posed we must have $\phi_1 = \phi_0^{\star}$. Otherwise, the problem
is well posed. From eq.(30) and eq.(31) we obtain
\begin{eqnarray}
Y^{\sigma + 2} &=& \mid \frac{1}{1 + X}\mid \\
{\rm where,} ~~~~~~~~~~~~~~~~~~Y &=& \mid \frac{\phi_1}{\phi_0} \mid \\
{\rm and}~~~~~~~~~~~~~~~~~~~~
X &=& \frac{\chi G_{0,0}^0(E) \mid \phi_0 \mid^{\sigma} (Y^{\sigma} -
1)}{1 + \chi G_{0,0}^0(E) \mid \phi_0 \mid^{\sigma}}
\end{eqnarray}
It is to be noted that for $\mid \phi_0 \mid = \mid \phi_1 \mid$,
eq.(36) becomes an identity. Consider now the case where $\chi
G_{0,0}^0(E) > 0$. This limit can be obtained either by (i) $\chi > 0,
E > 2$ or by (ii) $\chi < 0, E < -2$. It is easy to see that if $Y
\ne 1$, eq.(36) is not satisfied. So, the solution in this
limit is $\mid \phi_0 \mid = \mid \phi_1 \mid$. Since these
quantities can be taken real, we have $\phi_0 = \pm \phi_1$. Consider
next the other limit where $\chi G_{0,0}^0(E) < 0$. To achieve this
limit we need either (i) $\chi < 0, E > 2$, or (ii) $\chi > 0, E <
-2$. Now it can be easily shown that if both $D_0 = 1 - \mid \chi
G_{0,0}^0(E) \mid \mid \phi_0\mid^{\sigma} < 0$ and $D_1 = 1 - \mid
\chi G_{0,0}^0(E) \mid \mid \phi_1 \mid^{\sigma} < 0$ the solution
of eq.(36), is again $\mid \phi_0 \mid = \mid \phi _1 \mid$.
Note further that if $Y > 1$ is a solution of eq.(36), $Y < 1$ should
also be a solution of eq.(36) [see for example eq.(35)]. 
So, any constraint on $D_0$ will
imply similar constraint on $D_1$ and vice-versa. On the other hand we
may have $\mid \phi_0 \mid \ne \mid \phi_1 \mid $ if $D_0 > 0$ and
consequently also $D_1 > 0$. We consider again eq.(30) which
for $\chi G_{0,0}^0(E) < 0$ is
\begin{equation}
1 - \mid \chi G_{0,0}^0(E) \mid \mid \phi_0 \mid^{\sigma} = \mid \chi
G_{0,0}^0(E) \mid \mid \phi_1 \mid^{\sigma} \eta ~~sgn(E)
(\frac{\phi_1}{\phi_0 })
\end{equation}
We can also consider the equation for $\phi_1$ (eq.(31)). Both 
equations will yield the same conclusion. It is easy to see that 
for $E > 2$, eq.(39) tells that $(\phi_1 / \phi_0)$ must be 
positive. On the other hand for $E < -2$, $(\phi_1 / \phi_0)$ 
should be negative. This result coupled with the formula for 
${\phi_n}$ (eq.(32) and eq.(33)) suggests that SL states 
for $ D_0 > 0$, either have no nodes for $E > 2$ or have N 
(the number of sites in the lattice) nodes
for $E < -2$. We further note that band edge states in the perfect
system have this characteristics. In case of the linear eigenvalue
problem, we know that there will be only one SL state in this
limit. This SL state will evolve from one of the band edge
states. Since here we are all dealing with a nonlinear eigenvalue
problem it is in principle possible to have multiple permissible
values of $\phi_0$ and $\phi_1$. In other words, we may get a set of
states. We shall refer to this a symmetric set. But all these 
SL states will contain same number of nodes. Of course, a 
subset of these states will have $\phi_0 = \phi_1$ and one state 
in this subset will survive for $\sigma = 0$.

In the linear dimer impurity problem we also get another SL
state. This state has either a node ($E > 2$) through the dimer or 
the node through the dimer disappears ($E < -2$). We should have the 
same situation with the nonlinear dimer, but instead of having one 
such state we may have a set of states again with the same number of 
nodes. Since in the second set we must have $\phi_1 / \phi_0 = - sgn(E)
\mid (\phi_1 / \phi_0) \mid$, for this set to appear we need $D_0,D_1 < 0$.
This, in turn implies that the second set of states should always
have $\mid \phi_0 \mid = \mid \phi_1 \mid$. So, states where $\phi_0 
= \phi_1$ in the first set together with states coming from this 
antisymmetric set will constitute a major part of the
phase diagram for SL states. This phase diagram will be
discussed here. We also analyze if $\phi_0 \ne \phi_1$ is at all
possible in the first set of SL states.

For $\mid \phi_0 \mid = \mid \phi_1 \mid$ from either eq.(30) or
eq.(31) we obtain for $\mid E \mid > 2,$
\begin{equation}
\eta_{\pm} = \frac{1}{\mid \chi \mid \mid \phi_0 \mid^{\sigma} \pm 1}
\end{equation}
For positive sign we need $\frac{\phi_1}{\phi_0} = -sgn(\chi) = 
sgn(E)$ while for the other sign $\frac{\phi_1}{\phi_0} =  
sgn(\chi) = - sgn(E)$. Furthermore, the equation for determining 
$\mid \phi_0 \mid$ can be obtained from eq.(34) and eq.(40). 
The required equation for $\mid \phi_0 \mid$ is
\begin{equation}
(1 - 2 \mid \phi_0 \mid^2) ~(\chi \mid \phi_0 \mid^{\sigma} 
\pm 1)^2 = 1
\end{equation}
and the energy of the SL state ($E$) is $E = sgn(E) (\eta +
\frac{1}{\eta}),$ $0 \le \eta \le 1$.

The phase diagram is obtained by numerical calculation (see Fig.4).
Note that for $\sigma = 0$, we have one state for $\mid \chi \mid <
2$ and then for $\mid \chi \mid > 2$ second state appears. 
This agrees with established result \cite{46}. On the other hand, 
we find that for $0 \le \sigma \le 2$ we find a regime (I) with one 
SL state (coming only from symmetric set). For all values 
of $\sigma$ in this range there is a $\mid \chi_{cr} \mid$ which increases 
with increasing $\sigma$, shown in figure by dotted line. Two 
SL states exist along this line. Note further that one of 
these states coming from the antisymmetric set. Above this line we 
then have a regime (IV) where three SL states are obtained ( one
from the symmetric set and other two from the antisymmetric set ).

We now consider the case where $\sigma > 2$. We have a region (II) 
where no SL state exists. Note that similar situation is also
obtained for one nonlinear impurity case \cite{41}. This region is bounded
by a critical line along which one SL state exists ( shown by solid
line in Fig.4). This state comes from symmetric set.
Above this critical line we have a region (III) of two SL states.
Furthermore, we never get more than two states from the symmetric
set. This region of two states continues until we reach the critical
line ( shown by dotted line in the Fig.4 ) due to the 
antisymmetric set. Along this line we always have
three SL states and above this line we have a region (V) of four
SL states. We also mention that $\sigma = 2$ is the special
line. Along this line all transitions are taking place. For example,
$\mid \chi_{cr} \mid$ at $\sigma = 2$ below which no SL state is
obtained, is 1. This has been contrasted with the published result of
$\mid \chi_{cr} \mid \sim 2.0$ \cite{40}. Another $\mid \chi_{cr} \mid$
is found to be
exactly 8 which come from the antisymmetric set. We also emphasize 
that $\sigma = 2$ line is a line of continuity. The energy of states 
as a function $\chi$ in regions I and IV of Fig.4 is shown in 
Fig.5. we have taken $\sigma = 1$ for example to analyze this. 
In region I the energy of the state always increases with 
increasing $\mid \chi \mid$. In region IV, the energy of one of the 
two states belonging to the antisymmetric set decreases with  
increasing $\mid \chi \mid$ and it is asymptotically reducing to 
$E = 2$ ( band edge state ). The energy diagram in regions II, 
III and V is shown in Fig.6. In this case we have taken $\sigma = 3$.
The energy diagram also shows three distinct regions having no state, 
two states and four states. Two consecutive regions are separated 
by critical lines where one and three states exist respectively. 
We should note that the energy of one pair of states ( one from 
the symmetric set and the other from the antisymmetric set ) 
decreases asymptotically to $E = 2$ ( band edge state ) with 
increasing $\mid \chi \mid$. The energy of other pair increases 
as $\mid \chi \mid$ increases.

We now provide the analytical support for our numerical results. Since we
are considering the case $\mid \phi_0 \mid^2 = \mid \phi_1 \mid^2$ 
here, from eq.(34) we get $\mid \phi_0 \mid^2 = (1 - \eta^2)/2$. 
Then, from eq.(40) we obtain
\begin{equation}
\frac{2^{\sigma / 2}}{\mid \chi \mid} = \eta_{\pm} (1 \mp
\eta_{\pm})^{-1} (1 - \eta_{\pm}^2)^{\sigma /2}
\end{equation}
\noindent In eq.(42) the upper case sign is for symmetric set while
the lower case sign represents the antisymmetric set. We consider
first the symmetric case. We denote the right hand side of eq.(42) by
$g_s(\sigma, \eta)$
\begin{equation}
g_{s}(\sigma, \eta) = \eta (1 + \eta)^{\frac{\sigma}{2}} (1 -
\eta)^{\frac{\sigma}{2} -1}
\end{equation}
and $\eta$ lies in [0, 1]. We further note that for $\sigma < 2$, the
domain of $g_{s}(\sigma, \eta)$ is $[0, \infty]$. So, by applying our
analysis of one nonlinear impurity in one dimensional system we
obtain that for $\sigma < 2$, the system will produce one and only
one SL state for $\mid \chi_{cr} \mid > 0$. In other words, in
this limit $\mid \chi \mid = 0$. Precisely this is seen in our
numerical calculation. On the other hand, for $\sigma > 2,
~g_{s}(\sigma, 0) = 0 = g_{s}(\sigma, 1)$. Furthermore,
$g_{s}(\sigma, \eta)$ is positive semidefinite for $\eta$ in [0, 1].
This, in turn implies that there will be a regime where no SL
state will exist. However, for $\mid \chi \mid$ exceeding
some critical value, $\mid \chi_{cr} \mid$, we shall obtain at least
two SL states. It is quite transparent from the form of
$g_{s}(\sigma, \eta)$ that the function has only one maximum with
respect to $\eta$ in [0, 1]. Our numerical calculations also suggest
so. Therefore, there will be a no SL state region and a two
localizes states region separated by a line in the $(\chi, \sigma)$
plane. Furthermore, only one SL state will exist along this
line. To obtain the equation for this line we as usual maximize
$g_{s}(\sigma, \eta)$ with respect to $\eta$. This then yields
\begin{equation}
\sigma \eta^2 - \eta - 1 = 0
\end{equation}
and the permissible solution for $\eta, ~\eta_{ms}$ is 
\begin{equation}
\eta_{ms} = \frac{1 + \sqrt{1 + 4 \sigma}}{2 \sigma}
\end{equation}
Now from eq.(43) we obtain for $\sigma > 2$
\begin{equation}
\mid \chi_{cr} \mid = \frac{2^{\sigma /2}}{g_{s}(\sigma, ~\eta_{ms})}
\end{equation}
\noindent For $\sigma = 2, ~g_{s}(2, ~\eta) = 2$ is the maximum
permissible value of $g_{s}(2, ~\eta)$ for $\eta$ in [0, 1]. Albeit
this is not a true maximum of $g_{s}(\sigma, ~\eta)$, the left hand
side of eq.(42) must attend this value to obtain a SL state.
Hence, for $\sigma = 2, ~\mid \chi_{cr} \mid = 2$. Analytical
continuation of eq.(46) also yields this result. 

For the antisymmetric case the relevant function to analyze is 
\begin{equation}
g_{a}(\sigma, ~\eta) = \eta (1 - \eta )^{\frac{\sigma}{2}} (1 +
\eta)^{\frac{\sigma}{2} - 1}
\end{equation}
\noindent We note that for $\sigma > 0, ~g_{a}(\sigma, ~0) = 0 =
g_{a}(\sigma,~1)$. The function is also positive semidefinite for
$\eta$ in [0, 1]. Hence, there will be a maximum of $g_{(\sigma,
~\eta)}$ with respect to $\eta$ for $\eta$ in [0, 1]. It is obvious
from the structure of $g_{a}(\sigma, ~\eta)$ that there is only one
maximum. Form these we conclude that there will be two regions, one
having no SL state and the other containing two SL
states. Again the line separating these regions will contain only one
SL state. To obtain the equation of this line we maximize
$g_{a}(\sigma, ~\eta)$ with respect to $\eta$. We thus obtain
\begin{equation}
\sigma \eta^2 + \eta - 1 = 0 
\end{equation}
\noindent and the permissible solution of $\eta, ~\eta_{ma}$ is 
\begin{equation}
\eta_{ma} = \frac{\sqrt{4 \sigma + 1} - 1}{2 ~\sigma}
\end{equation}
\noindent So, the equation for the $\chi_{cr}$ line is
\begin{equation}
\mid \chi_{cr} \mid = \frac{2^{(\sigma /2)}}{g_{a}(\sigma, 
~\eta_{ma})}
\end{equation}
We again note that for $\sigma = 2, \eta_{am} = \frac{1}{2}$ 
which in turn yields $g_{a}(\sigma, \eta_{am}) =\frac{1}{4}$. 
Hence, from equation.(47) we get $\mid \chi_{cr} \mid =
8$. This is obtained from out numerical calculation. For $\sigma =
0$, the maximum permissible value of $g_{a}(0, \eta)$ is
$\frac{1}{2}$ which occurs at $\eta = 1$. Consequently from eq.(50) 
we obtain $\mid \chi_{cr} \mid = 2$. This also agrees with the 
established result for static dimer impurities \cite{46}. We further 
note that the analytical continuation of eq.(50) also produce this 
result. We now note the followings. (1) If $\sigma \rightarrow
\infty$, $\eta_{ms} \sim \eta_{ma} = \frac{1}{\sqrt{\sigma}}$. This 
implies that two critical lines will approach each other as $\sigma$ 
increases. (2) In contrast to the symmetric case, the antisymmetric 
case will have nonzero $\mid \chi_{cr} \mid$ for all $\sigma ( > 0 )$
separating the no SL state and two SL states regimes. (3) The value of 
$\mid \chi_{cr} \mid$ for the antisymmetric case is greater than the 
corresponding value in the symmetric case if $\sigma$ is finite.
(4) For the antisymmetric case, there is no regime with one SL
state except the critical line. This is again in contrast to the
symmetric case but quite similar to the Caley tree case.

We now consider the possibility of $\phi_0 \ne \phi_1$ in the region
where $D_0$, $D_1 > 0$. Since $E > 2$, $-\chi > 0$, we are taking
positive $\chi$ strictly to avoid complication. We obtain from
equation for $\phi_n$, $n \in N$
\begin{eqnarray}
\chi (\mid \phi_0 \mid^{\sigma} - \mid \phi_1 \mid^{\sigma}) &=&
\frac{\phi_0^2 - \phi_1^2}{\phi_1 \phi_0} \\
{\rm and}~~~~~~~~~~~~~~~~~~
\eta = \frac{1}{\chi \mid \phi_0 \mid^{\sigma} +
\frac{\phi_1}{\phi_0}} &=& \frac{1}{\chi \mid \phi_1 \mid^{\sigma} +
\frac{\phi_0}{\phi_1}}
\end{eqnarray}
Furthermore, we have eq.(34). We emphasize that not only that
$\phi_0$ and $\phi_1$ have same sign, they are real quantities. For
$\sigma = 2$, we find after some algebra that 
\begin{eqnarray}
\eta^3 &-& \eta + \frac{1}{\chi} = 0 \\
{\rm and}~~~~~~~~~~~~~~~~~~ \beta &=& \frac{\phi_1}{\phi_0} = 
\frac{1 \pm \sqrt{1 - 4 \eta^2}}{2 \eta} 
\end{eqnarray}
\noindent Since $\beta$ should be real, we must have $\eta <
\frac{1}{2}$, Then from eq.(53) we find $\mid \chi_{cr} \mid =
\frac{8}{3}$ for this case. This is shown by
a box in Fig.4. We emphasize that to the best of our
knowledge this result has not been obtained before.  
So, along the $\sigma = 2$ line we have (i) $\mid \chi \mid < 1$, 
no SL state, (ii) $1 \le \mid \chi \mid \le \frac{8}{3}$, one 
SL state, (iii) $\frac{8}{3} < \mid \chi \mid < 8$, two SL state, 
(iv) $\mid \chi \mid$ = 8, a point of three SL states and (v) $\mid 
\chi \mid > 8$, four SL states. The case of $\sigma$ = 1, 3 and 
4 have also been examined. No SL state for $\phi_0 \ne \phi_1$ 
is found. It appears that $\sigma = 2$ is a very special case. 
Further analysis for arbitrary $\sigma$ will be presented elsewhere.

\section{A rotational nonlinear dimer impurity in a perfect chain}

We consider now the case where $f_{m}(\mid C_{m} \mid) = \mid C_{m}
\mid^2 /(\sqrt{1 + (\frac{\chi}{\Delta})^2 \mid C_m \mid^4})$ for $m$
= 0 and 1 and zero otherwise. In other words we have a nonlinear
dimer impurities made up of so called rotational nonlinear
impurities \cite{45}. In this case also $\phi_0 = \pm \phi_1$
constitute a set of solutions. We focus our attention on this set. 
More general case will be considered elsewhere. We have here two 
cases, namely the symmetric case, where $\phi_0 = \phi_1$ and the 
antisymmetric case otherwise. For the symmetric case from the 
appropriate version of eq.(40) along with eq.(34) we obtain
\begin{equation}
\frac{4 \Delta^2}{\chi^2} = (1 + \eta)^2 [\Delta^2 \eta^2 - (1 -
\eta)^2] = G_{\Delta}^{+}(\eta)
\end{equation}
\noindent Since the left hand side of eq.(55) is always positive, we
need $\eta \ge 1/(1 + \Delta)$ but at the same time we need $\eta \le
1$. So, the condition on $\eta$ is $1/(1 + \Delta) \le \eta \le 1$.
It is obvious that the constraint is met for $\Delta \ge 0$.
On the other hand, $G_{\Delta}^{+}(1) = 4 \Delta^2$.
Since, $G_{\Delta}^{+}(\eta)$ is monotonically increasing
function of $\eta$ for $\eta$ in [0, 1], $4 \Delta^2$ is the maximum
permissible value of the function. So, for a SL state to appear we need
$\mid \chi_{cr} \mid = 1$. Since, there is only one intersection of a
line $y = \frac{4 \Delta^2}{\chi^2}$ = constant with the curve
$G_{\Delta}^{+}(\eta)$ for $\eta$ in [0, 1] and $\mid \chi \mid > 1$,
we shall always get one and only one SL state from
the symmetric case if $\mid \chi \mid \ge \mid \chi_{cr} \mid = 1$.

For the antisymmetric case from the appropriate version of eq.(40)
and eq.(34) we obtain
\begin{equation}
\frac{4 \Delta^2}{\chi^2} = (1 - \eta)^2 [\Delta^2 \eta^2 - (1 +
\eta)^2] = G_{\Delta}^{-}(\eta)
\end{equation}
Since $G_{\Delta}^{-}(\eta)$ should be positive semidefinite in the
permissible interval of $\eta$ not exceeding [0, 1], we need
$\eta_{min} \ge \frac{1}{\Delta - 1}$. Since for $\eta \le 1$, we
must have $\Delta \ge 2$. We note now that $G_{\Delta}^{-}(1) = 0 =
G_{\Delta}^{-}(\eta_{min})$. So, there must be at least one maximum
of $G_{\Delta}^{-}(\eta)$ with respect to $\eta$. Imposing the
condition of maximality on $G_{\Delta}^{-}(\eta)$ with respect to
$\eta$ we get
\begin{equation}
\eta_{max} = \frac{\Delta^2 + 2}{2 (\Delta^2 - 1)}
\end{equation}
It can be easily shown that $\eta_{max} \le 1$. Furthermore $\eta_
{max} \not< \eta_{min}$. Then by setting $\eta_{max} =
\eta_{min} = \frac{1}{\Delta -1}$, we again find that $\Delta_{cr}$ 
= 2. We also note that $G_{\Delta}^{-}(\eta)$ has only one allowed 
maximum for $\eta$ in $[\eta_{min}, 1]$. Hence, above this maximum 
we shall not obtain any SL state. On the other hand for $\frac{4
\Delta^2}{\chi^2}$ lying below this maximum, we shall get two
SL states. The equation of the line separating these two 
regions is 
\begin{equation}
\mid \chi_{cr} \mid = \frac{2 \Delta}{\sqrt{G_{\Delta}^{-}(\eta_{max})}}
\end{equation}
We further note that eq.(56) and eq.(57) in the limit $\Delta 
\rightarrow \infty$ yields $\mid \chi_{cr} \mid = 8$. This is expected. 

In Fig.7 we have shown graphically the existence or nonexistence of
SL states for the antisymmetric case with $\Delta = 4$ for example. 
The plot of $G_{\Delta}^- (\eta)$ is shown as dotted curve. Along 
with that we have superposed the lines $y = \frac{4 \Delta^2}{\chi
^2}$ for three values of $\chi$ as shown in figure. The line with 
$\mid \chi \mid = 11.313$ touches the maximum point of the dotted 
curve and represents one SL state. So any line with $\mid \chi \mid 
< 11.313$ will not touch the dotted curve and hence there are no SL 
states for $\mid \chi \mid < 11.313$ ( for example the upper line 
is for $\mid \chi \mid = 9.65$ ). On the other hand the lines with 
$\mid \chi \mid > 11.313$ will touch the line twice.
Therefore there will exists two SL states for $\mid \chi \mid > 
11.313$ ( for example the third line with $\mid \chi \mid = 14.61$ ).
Therefore it is very transparent that for this case with $\Delta 
= 4, \mid \chi_{cr} \mid = 11.313$.

We consider next the full phase diagram obtained numerically. This is
shown in Fig.8. For $\mid \chi \mid < 1$ we
have a no state region for any $\Delta$. For $\mid \chi \mid > 1$,
and $\Delta \le 2$, we have a region where one SL state
exists ( coming from symmetric set ). For $\mid \chi \mid > 1$ 
and $\Delta > 2$, we have two regions separated by critical line (shown
by solid line) 
in the $(\chi, \Delta)$ plane. Above this critical line we have 
three SL states ( one from symmetric set and other
two from antisymmetric set ) but below this line we have only one 
state ( coming from symmetric set ). On the critical line
only two SL states exists ( one from symmetric set and the
other one from antisymmetric set ). Energy of SL states as a
function of $\chi$ for $\Delta = 4$ is shown in Fig.9. It shows the
typical behavior.

\section{A model derivation of the DNLSE with generalized
nonlinearities} 

To start with we consider a model potential, $V(x)$ given by
\begin{equation}
V(x) = k [\frac{1 - cos(2\gamma x)}{4 \gamma^2}]^d
\end{equation}
When d = 1, we obtain the potential for the nonlinear pendulum. We
also know that such a system exhibits two types of motion, namely,
libration and rotation. Two regions are separated by a critical line,
called separatrix \cite{47}. Similar situation also occurs for general d.
When d = 1/2, the periodic part of the potential is saw--tooth type
with kinks smoothened out. For d $>>$ 1, the periodic part is more like
repeated hard wall potential. These are shown in Fig.10 We
further note that for $\gamma \rightarrow 0$, V(x) $\approx
\frac{k}{2^d} \mid x \mid^{2d}$. For d = 1, we get the usual Hook
spring. On the other hand for general d the spring deviates from the
Hookian behavior.

We consider next a Hamiltonian, $H$ given by
\begin{equation}
H = \frac{1}{2} \sum_{n=1}^{N} [\omega_n(x_n) \mid C_n \mid^2 + (C_n
C_{n+1}^{\star} + C_n^{\star} C_{n+1})] + \sum_{n=1}^{N}
[\frac{p_n^2}{2m_n} + V_n(x_n)] = H_1 + H_2
\end{equation} 
\noindent By defining positions $(Q_n)$ and momenta $(P_n)$ by $Q_n =
(C_n + C_n^{\star})/2$ and $P_n = i (C_n^{\star} - C_n)/2$
respectively, we find that 
\begin{equation}
H_1 = \frac{1}{2} \sum_n \omega_n(x_n) (P_n^2 + Q_n^2) + \sum_n (P_n
P_{n+1} + Q_n Q_{n+1})
\end{equation}
So, $H_1$ can be said to define the dynamics of a super oscillator with
2N degrees of freedom. Furthermore, the frequency of the oscillator
in a given direction is coupled to the motion of another oscillator
defined by $x_n$ \cite{7}. On the other hand, we can replace $C_n
(C_n^{\star})$ in $H_1$ by the annihilation (creation) operator $C_n
(C_n^{\dagger})$ such that $[C_n^{\dagger}, C_n] = i \hbar$. $C_n
(C_n^{\dagger})$ then defines the annihilation (creation) operator of a
particle. Under this transformation it transforms to the usual tight
binding Hamiltonian of a particle moving on a ring of N sites.
Parameters {$x_n$} then define the displacement of lattice points
from their respective equilibrium positions. Although we are
considering here the local displacement of atoms ( optical phonons ),
the collective motion of atoms in the lattice ( acoustic phonon ) can
be incorporated in this formalism by making appropriate modifications
of $H$ \cite{48}. Now from Hamiltons principle we find that
\begin{eqnarray}
i \dot C_m &=& 2 \frac{\partial{H}}{\partial{C_m^{\star}}} =
\omega_m(x_m) + C_{m+1} + C_{m-1} \\
\dot x_l &=& \frac{\partial{H}}{\partial{p_l}} = \frac{p_l}{m_l} \\
{\rm and}~~~~~~~~~~~~~~~-\dot p_l &=& \frac{\partial{H}}{\partial{x_l}}
= \omega_l^{'}(x_l) \mid C_l \mid^2 + V_l^{'}(x_l)
\end{eqnarray}
\noindent We now assume that the dynamical evolution of {$x_l$}
occurs in a much longer time scale compared to the time scale of
{$C_l$}. So, the dynamics of the other system is determined primarily
by the adiabatic variation of {$x_n$} \cite{7}. In the electron lattice
interaction language, the dynamics of the lattice is assumed to be
far far slow compared to the dynamics of the electron. Since the
adiabatic approximation for {$x_l$} implies $\dot p_l = 0$ for $l \in
N$, the equation for determining ${x_l}$ is then 
\begin{equation}
\omega_l^{'}(x_l) \mid C_l \mid^2 = - V_l^{'}(x_l)
\end{equation}
\noindent Consider, for example, $V_l(x_l) = \frac{1}{2} k_l x_l^2$
and $\omega_l(x_l) = E_l x_l$. Then from eq.(65) we obtain $x_l =
(E_l / k_l) \mid C_l \mid^2$. Introduction of this in eq.(62) yields
the DNLSE where $f_l(\mid C_l \mid) = \mid C_l \mid^2$ and $\chi_l =
E_l^2 / k_l$.

In principle $\omega_n(x_n)$ can be nonlinear function of $x_n$.
Furthermore, the overall dynamics of {$x_n$} can be quite complicated
and yet can be physically realistic. A few example in this regard
have already been considered in the literature \cite{41,42,43,45}. 
We consider a situation where for $n \in N$ $V_n(x_n)$ is given by 
eq.(59) and for $\omega(x)$ we assume 
\begin{equation}
\omega(x) = \frac{E_0}{\gamma} \left(\frac{1 - cos(2 \gamma x)}{4
\gamma^2}\right)^{d-1} sin(2 \gamma x)
\end{equation}
\noindent Then from eq.(65) we get
\begin{eqnarray}
[1 - \delta^2 (\frac{d - 1}{d})^2 \mid C \mid^4] tan^2(2 \gamma x) -
2 \delta \mid C \mid^2 tan(2 \gamma x) + \delta^2 (\frac{2d -
1}{d^2}) \mid C \mid^4 = 0 \\
{\rm where,}~~~~~~~~~~~~\delta = \frac{4 E_0 \gamma}{k} =
\frac{4 E_0^2 / k}{E_0 / \gamma} = \frac{\chi}{\Delta}~~~~~~~~~~~~~~~
\end{eqnarray}
\noindent Eq.(67) then yields
\begin{eqnarray}
tan(2 \gamma x) &=& \frac{\delta \mid C \mid^2}{y} [1 + \sqrt{1 - y
\frac{(2d - 1)}{d^2}}] \\
{\rm where,}~~~~~~~~~~~~~~~~y = 1 &-& \frac{\chi^2}{\Delta^2} (\frac{d
- 1}{d})^2 \mid C \mid^4
\end{eqnarray}
\noindent Then for d = 1 we have y = 1. This in turn yields $tan(2
\gamma x) = \delta \mid C \mid^2$. Then after some simple algebra we
get 
\begin{equation}
\omega(x) = -\frac{\chi \mid C \mid^2}{\sqrt{1 +
(\frac{\chi}{\Delta})^2 \mid C \mid^4}}
\end{equation}
\noindent We consider next the other limit where $\gamma \rightarrow
0$. Then from eq.(69) we obtain
\begin{equation}
x \approx \frac{2 E_0}{k} (2d - 1) \mid C \mid^2
\end{equation}
\noindent This in turn yields $\omega(x) = - \chi_{\sigma} \mid C
\mid^{\sigma}$ where $\sigma = 2 (2d - 1)$ and
\begin{equation}
\chi_{\sigma} = -[2^{d+2} \mid E_0 \mid (\frac{\mid E_0 \mid (2d -
1)}{k})^{2d - 1}]
\end{equation}
\noindent Then for d = 1, we get $\omega(x) = \chi \mid C \mid^2$.
Introducing this in eq.(62) we get the desired DNLSE.

We consider another model where we take $\omega(x) = 2 E_0 x$ and
$V(x)$ is given in eq.(59). Then again from eq.(65) we get 
\begin{equation}
E_0 \mid C \mid^2 = - k d \mid x \mid^{2(d - 1)} x
\end{equation}
\noindent This in turn yields
\begin{equation}
\mid x \mid = \mid \frac{E_0}{kd}\mid^{\frac{1}{2d-1}} \mid C
\mid^{\frac{2}{2d -1}}
\end{equation}
\noindent Consequently we have $\omega(x) = - \widetilde
\chi(\sigma^{'}) \mid C \mid^{\sigma}$ where, $\sigma^{'} = 2/(2d-1)$ and
\begin{equation}
\widetilde \chi(\sigma^{'}) = 2 \mid E_0 \mid \frac{E_0}{kd}
\mid^{\frac{1}{2d-1}}
\end{equation}
\noindent Then inserting the $\omega(x)$ in eq.(62) we obtain the
desired set of equation.

\section{Conclusion}

The formation of stationary states due to a single nonlinear impurity
in a Caley tree and a dimeric
nonlinear impurity in a perfect 1D linear system has been studied here
using the DNLSE. Two types of nonlinear impurities, namely, $f(\mid C
\mid) = \mid C \mid^{\sigma}$ where $\sigma$ is arbitrary and
the rotational impurity are considered. Altogether four cases are
studied. Important features of these problems are thoroughly
discussed in the text. Whenever necessary analytical
arguments in support of numerical results are provided. Some
important aspects of this paper are elucidated further below.

In this paper a very useful synthesis of the usual Greens function
approach and the ansatz approach is made to gain a better
understanding of the problems. Because of synthesis we are able to
derive for the first time many important results. For example, in the
case of nonlinear dimer with $f(\mid C \mid) = \mid C \mid^{\sigma}$,
at $\sigma = 2$ and $\mid \chi \mid > \frac{8}{3}$ a stationary
localized state in which $\mid \phi_0 \mid \ne \mid \phi_1 \mid$ is
obtained. However, for $\sigma = 1, 3$ and 4, no such $\chi_{cr}$ is
found. So, it appears that for this nonlinear
dimer, $\sigma = 2$ is a very special case. Of course, further
analysis is required and this will be presented elsewhere.

For the nonlinear dimer problem, $\mid \chi_{cr} \mid$ $\sim 2$ and 
$3.6$ have been reported in the literature. These values of $\chi_{cr}$
separate the whole of the $\sigma = 2$ line into three regions,
having no SL state, two SL states and three SL states respectively.
This has to be contrasted with our analysis giving three values of
$\mid \chi_{cr} \mid$, namely, 1, $\frac{8}{3}$ and 8. So, we have 
a new $\mid \chi_{cr} \mid$ at 8 separating two SL states and four SL
states regions with three SL states at that point. Furthermore, 
our other two values of $\mid \chi_{cr} \mid$ are lower than the 
reported values.

For the monomer problem $\mid \chi_{cr} \mid$ = 8 does not
exist. However, one obtains $\mid \chi_{cr} \mid = 2$ which separates
the no SL state and two SL states regions. The analytical reason
behind this is discussed here. Furthermore, the equation for the
critical line separating these regimes is analytically derived. In
fact, equations describing various critical lines are derived from
general analytical approach. Our numerical results also agree very well
with our analytical results. We further note that for the dimer
problem this $\mid \chi_{cr} \mid$ is 1. So, it appears that for a
cluster of N sites, this $\mid \chi_{cr} \mid$ will go as
$\frac{1}{N}$. This assertion, however, needs thorough investigation.

Another interesting feature of this problem is the possibility of
more localized states than the number of impurities in certain
situations. This happens due to multiple permissible values of the
amplitude at the impurity sites. Furthermore, the localization length
of one set of states increases while this length in the other set of
states decreases as the strength of the nonlinearity parameter,
($\chi$) increases. Hence, in contrast to the static impurity case,
the system here cam assume more than one configuration.

Physically, the effect of localized states is manifested in transport
properties of the material. However, a system containing a finite
number of impurities can only produce finite number of such states.
So, the probability that these states will exert their influence on
transport properties of the system is very small, if not unlikely. On
the other hand for a thorough understanding of properties of
disordered system comprising of nonlinear impurities and perfect
sites, a critical understanding of these problems is essential. So,
the importance of the problems presented here should be understood in
this context.

\begin{figure}
\caption {Critical value of nonlinearity ($\chi_{cr}$) separating the
`No SL state region' from the `Two SL states region' is plotted as a 
function of $\sigma$ for Caley trees having K = 1 (solid line), K = 2 
(dashed line), K = 3 (dotted line) and K = 10 (dot-dashed line).
There is one nonlinear impurity in the Caley tree of the form -$\chi
\mid C \mid^{\sigma}$}

\caption {The energy of SL states shown in Fig. 1 for $\sigma = 1$ 
is plotted as a function of the nonlinearity parameter ($\chi$) for 
the Caley tree. The solid curve is for K = 1, the dotted curve is 
for K = 2, the dashed curve is for K = 3 and the dot-dashed curve 
is for K = 10.}

\caption {The phase diagram for the Caley tree with a single rotational
nonlinear impurity showing the number of SL states as a function of
nonlinearity parameter ($\chi$) and the parameter $\Delta$. In this case
the connectivity of the Caley tree, K = 2.}

\caption {The phase diagram for a perfect chain with a nonlinear dimer
made up of impurities of the form -$\chi \mid C \mid^{\sigma}$. 
The figure shows regions containing different number of SL
states in the ($\chi, \sigma$) plane. There are one SL state in 
region I, no SL state in region II, two SL states in region III, 
three SL states in region IV and four SL states in region V. The 
solid $\mid \chi_{cr} \mid$ curve arises from the symmetric set 
and the dotted curve arises from the antisymmetric set. At $\sigma$ 
= 2 three values of $\mid \chi_{cr} \mid$ exist. These points are 
shown by boxes. They are 1, 8/3 and 8 respectively.}

\caption {The energy diagram of SL states as a function of $\chi$ at 
$\sigma$ = 1 which corresponds to the regions I and IV of Fig.4. The
solid curve is due to the symmetric set and the dotted curve is due to
the antisymmetric set.}

\caption {The energy diagram for SL states as a function of $\chi$ at
$\sigma$ = 3 which corresponds to the regions II, III, V in Fig.4.
The solid curve is due to the symmetric set and the dotted curve is due
to the antisymmetric set. The dashed vertical lines touch the solid
and dotted curves at the critical values of $\mid \chi \mid$.}

\caption {The function $G_{\Delta}^-(\eta)$ is plotted as a function 
of $\eta$ (dotted curve). The solid lines correspond to the y = $\frac{4
\Delta^2}{\chi^2}$ for $\mid \chi \mid$ = 9.65, 11.313 and 14.61 
respectively. Here $\Delta = 4$}

\caption {The phase diagram for the perfect chain with a rotational
nonlinear dimer impurity showing the number of SL states as a
function of $\chi$ and $\Delta$.}

\caption {The energy for SL states shown in Fig. 8  is plotted as 
a function of $\chi$ for $\Delta = 4$.}

\caption {The potential $V(x) = k (\frac{1 - cos(2 \gamma x)}{4
\gamma^2})^{\sigma}$ is plotted as a function of $x$. The solid curve
is for $k = 1, \gamma = 1$, $\sigma = 0.5$ and the dotted one is for
$k = 1, \gamma = 1$ and $\sigma = 2$.}
\end{figure}

\begin{thebibliography}{99}
\bibitem{1} T. D. Holstein, Ann. Phys. (N. Y) {\bf 8} 325 (1959);
{\bf 8} 343 (1989).
\bibitem{2} A. S. Davydov, J. Theor. Biol. {\bf 38} 559 (1973);
Usp. Fiz. Nauk. {\bf 138} 603 (1982) and references therein.
\bibitem{3} A. C. Scott, F. Y. Chu and D. W. McLaughlin, Proc. I.
E. E. E. {\bf 61} 1443 (1973).
\bibitem{4} Eilbeck, P. S. Lomdahl and A. C. Scott, Physica D {\bf
16} 318 (1985).
\bibitem{5} J. M. Hyman, D. W. McLaughlin and A. C. Scott, Physica
D {\bf 3} 28 (1981).
\bibitem{6} D. W. Brown, K. Lindenberg and B. J. West, Phys. Rev.
B {\bf 37} 2946 (1988).
\bibitem{7} {\it Davydov's Soliton Revisited: Self trapping of
vibrational energy in protein, {\rm Vol. 243 of} NATO Advanced Study
Institute, Series B: Physics,}edited by Peter L. Christian and Alwyn 
C. Scott (Plenum, New York, 1991).
\bibitem{8} D. Chen, M. I. Molina and G. P. Tsironis, J. Phys.
Cond. Matt. {\bf 5} 8689 (1993).
\bibitem{9} Yi Wan and C. M. Soukoulis, Phys. Rev. B {\bf 40} 12264
(1989).
\bibitem{10} Yi Wan and C. M. Soukoulis, Phys. Rev. A {\bf 41} 800 (1990).
\bibitem{11} V. M. Kenkre, Physica D {\bf 68} 153 (1993).
\bibitem{12} M. I. Molina and G. P. Tsironis, Phys. Rev. Lett.
{\bf 73} 464 (1994).
\bibitem{13} M. Johansson, M. Hornquist and R.Riklund, 
Phys. Rev. B {\bf 52} 231 (1995).
\bibitem{14} V. M. Kenkre and D. K. Campbell, Phys. Rev. B {\bf 34}
4959 (1986).
\bibitem{15} V. M. Kenkre, G. P. Tsironis, Phys. Rev. B {\bf 35}
1473 (1987).
\bibitem{16} V. M. Kenkre, G. P. Tsironis and D. K. Campbell, {\it
Nonlinearity in Condensed Matter, ed.} A. R. Bishop, D. K. Campbell,
P. Kumar and S. E. Trullinger, (Springer-Verlag 1987).
\bibitem{17} G. P. Tsironis and V. M. Kenkre, Phys. Lett. A {\bf
127} 209 (1988).
\bibitem{18} V. M. Kenkre and G. P. Tsironis, Chem. Phys. {\bf 128}
219 (1988).
\bibitem{19} G. P. Tsironis, V. M. Kenkre and D. Finley, Phys. Rev.
A {\bf 37} 4474 (1988).
\bibitem{20} V. M. Kenkre and M. Kus, Phys. Rev. B {\bf 46} 13792
(1992).
\bibitem{21} V. M. Kenkre and M. Kus, Phys. Rev. B {\bf 49} 5956
(1994).
\bibitem{22} V. M. Kenkre and P. Grigolini, Z. Phys. B {\bf 90} 247
(1993).
\bibitem{23} V. M. Kenkre and H. -L. Wu, Phys. Rev. B {\bf 39} 6907
(1989). 
\bibitem{24} D. Hennig and B. Esser, Phys. Rev A {\bf 46} 4569
(1992). 
\bibitem{25} M. I. Molina and G. P. Tsironis, Physica D {\bf 65} 267 (1993).
\bibitem{26} P. Grigolini, H. -L. Wu and V. M. Kenkre, Phys. Rev. B
{\bf 40} 7045 (1989).
\bibitem{27} H. Wipf, A. Magerl, S. M. Shapiro, S. K. Satija and
W. Thimlinson, Phys. Rev. Lett. {\bf 46} 947 (1981).
\bibitem{28} A. Magerl, A. J. Dianoux, H. Wipf, K. Neumaior and I.
S. Anderson, Phys. Rev. Lett. {\bf 56} 159 (1986).
\bibitem{29} D. H. Dunlap, V. M. Kenkre and P. Reineker, Phys. Rev.
B {\bf 47} 14842 (1992).
\bibitem{30} B. C. Gupta and K. Kundu, I. J. P, (September, 1996).
\bibitem{31} P. K. Dutta and K. Kundu, Phys. Rev B {\bf 53} 1 (1996).
\bibitem{32} S. Takeno and S. HOMMA, J. Phys. Soc. Jap.
{\bf 60} 731 (1991). 
\bibitem{33} Y. S. Kivshar, Phys. Rev. B {\bf 47} 11167 (1993).
\bibitem{34} Y. S. Kivshar, Phys. Lett. A {\bf 161} 80 (1991).
\bibitem{35} D. Cai, A. R. Bishop and N. Gronbech-Jensen, Phys Rev.
Lett. {\bf 72} 591 (1994).
\bibitem{36} D. Hennig, K. O. Rasmussen, G. P. Tsironis and H.
Gabriel, Phys. Rev. E {\bf 52} R4628 (1995).
\bibitem{37} M. J. Ablowitz and J. M. Ladik, J. Math. Phys. 
{\bf 17} 1011 (1976).
\bibitem{38} R. Scharf and A. R. Bishop, Phys. Rev. A {\bf 43}
6535 (1991).
\bibitem{39} M. Johansson and R. Riklund, Phys. Rev. B {\bf
49} 6587 (1994).
\bibitem{40} M. I. Molina and G. P. Tsironis, Phys. Rev. B {\bf 47}
15330 (1993).
\bibitem{41} G. P. Tsironis, M. I. Molina and D. Hennig, Phys Rev.
E {\bf 50} 2365 (1994).
\bibitem{42} Y. Y. Yiu, K. M. Ng and P. M. Hui, Phys. Lett. A {\bf
200} 325 (1995).
\bibitem{43} Y. Y. Yiu, K. M. Ng and P. M. Hui, Preprint.
\bibitem{44} D. CASSI, Europhys. Lett. {\bf 9} 627 (1989).
\bibitem{45} V. M. Kenkre, H. -L. Wu and I. Howard, Phys. Rev. B
{\bf 51} 15841 (1995).
\bibitem{46} {\it Green's Function in Quantum Physics} by E. N.
Economou, Springer-Verlag, Berlin Heidelberg New York Tokyo. (1983).
\bibitem{47} {\it Classical Mechanics}, p-290, H. Goldstein, Addison-Wesley
(1950).  
\bibitem{48} G. Kalosakas, G. P. Tsironis and E. N. Economou, J. Phys.
Cond. Matt. {\bf 6} 7847 (1994).
\end{thebibliography}
\end{document}